\documentclass{SciPost}

\hypersetup{
    colorlinks,
    linkcolor={red!50!black},
    citecolor={blue!50!black},
    urlcolor={blue!80!black}
}

\usepackage[bitstream-charter]{mathdesign}
\DeclareSymbolFont{usualmathcal}{OMS}{cmsy}{m}{n}
\DeclareSymbolFontAlphabet{\mathcal}{usualmathcal}

\fancypagestyle{SPstyle}{
\fancyhf{}
\lhead{\colorbox{scipostblue}{\bf \color{white} ~SciPost Physics }}
\rhead{{\bf \color{scipostdeepblue} ~Submission }}

\fancyfoot[C]{{\thepage}}
}

\newcommand{\tmop}[1]{\mathrm{#1}}
\newcommand{\mathd}{\mathrm{d}}
\newcommand{\mathe}{\mathrm{e}}

\begin{document}

\begin{center}{\Large \textbf{\color{scipostdeepblue}{
Molecular sorting on a fluctuating membrane\\
}}}\end{center}

\begin{center}\textbf{
D. Andreghetti \textsuperscript{1,2},
L. Dall'Asta\textsuperscript{1,2,3},
A. Gamba\textsuperscript{1,2,3,$\star$},
I. Kolokolov\textsuperscript{4,5} and
V. Lebedev\textsuperscript{4,5}
}\end{center}

\begin{center}
{\bf 1} Institute of Condensed Matter Physics and Complex Systems,\\
Department of Applied Science and Technology,
Politecnico di Torino, \\
Corso Duca degli Abruzzi 24, 10129 Torino, Italy
\\
{\bf 2} Istituto Nazionale di Fisica Nucleare (INFN),
Via Pietro Giuria, 1, 
10125 Torino,
Italy 
\\
{\bf 3} Italian Institute for Genomic Medicine,
Candiolo Cancer Institute, \\
Fondazione del Piemonte per l'Oncologia (FPO), 
Candiolo, 10060 Torino, Italy
\\
{\bf 4} L.D. Landau Institute for Theoretical Physics, \\
142432, Moscow Region, Chernogolovka, Ak.~Semenova 1-A,
Russia
\\
{\bf 5} National Research University Higher School of Economics, \\
101000, Myasnitskaya 20, Moscow, Russia

$\star$ \href{mailto:andrea.gamba@polito.it}{\small andrea.gamba@polito.it}
\end{center}

\section*{\color{scipostdeepblue}{Abstract}}
\textbf{\boldmath{%
Molecular sorting in biological membranes is essential for proper cellular function. It also plays a crucial role in the budding of enveloped viruses
from host cells. We recently proposed that this process is
driven by phase separation, where the formation and growth of sorting domains
depend primarily on 
direct
intermolecular interactions.
In addition to these, Casimir-like forces---arising
from entropic effects in fluctuating membranes%
---may
also play a significant role in the molecular distillation process.
Here, using a combination of theoretical analysis and numerical
simulations, we explore how 
Casimir-like forces between rigid membrane inclusions
contribute to sorting, particularly in the biologically relevant regime where 
direct
intermolecular interactions are weak. Our results 
show that 
these 
forces 
enhance molecular distillation by reducing the critical radius for the formation of
new sorting domains and facilitating the capture of molecules within these domains. We identify the relative rigidity of the membrane and supermolecular domains
as a key parameter controlling molecular sorting efficiency, offering new insights into the physical principles underlying molecular sorting in biological systems.
}}

\vspace{\baselineskip}



\vspace{10pt}
\noindent\rule{\textwidth}{1pt}
\tableofcontents
\noindent\rule{\textwidth}{1pt}
\vspace{10pt}


\section{Introduction}

Molecular sorting is a vital process in eukaryotic cells, where proteins and other biomolecules are sorted and encapsulated into lipid vesicles for
targeted transport to specific subcellular locations. This distillation process occurs on lipid membranes,  such as the plasma membrane~{\cite{KR18}}, endosomes, the Golgi apparatus~{\cite{AB99}}, and
the endoplasmic reticulum~{\cite{ZPS+12}}, where biomolecules can bind and diffuse laterally. Due to
a variety of direct and indirect interactions, these molecules aggregate into domains with distinct chemical compositions. These domains
can induce membrane bending and fission~{\cite{SHS10,SJB08,CAB16,FS08a}},
ultimately forming separated submicron lipid vesicles that are transported to their designated subcellular sites by molecular 
motors. In this way, lipid membranes
act as natural molecular distillers, promoting intracellular order
and compartmentalization and counteracting the homogenizing effects of diffusion. Disruption of molecular sorting in living cells
is implicated in severe pathologies, including cancer~\cite{MN08,SH11b}. On the other end of the spectrum, analogous molecular sorting processes are exploited by enveloped viruses, such as HIV, SARS-CoV, and influenza, for their assembly and budding from host cells~\cite{PGS02,RL11,SL20,MS21},
further underscoring the practical relevance of understanding the physical mechanisms of molecular sorting.

We have recently proposed a simple model of molecular sorting as a phase-separation process. In this context, 
the efficiency of sorting is found to be
optimal at intermediate values of intermolecular attraction forces~\cite{ZVS+21,FPP+22,PFA+23}.
This theoretical prediction is consistent
with experiments on endocytic sorting in living cells under near-physiological conditions~\cite{ZVS+21},
and with measurements performed on photoactivated systems, where the strength of
intermolecular attraction can be directly controlled~\cite{DKW+21}.
The interpretation of molecular sorting as a phase-separation process
is also coherent with the observation that sorting domains in living cells exhibit a critical size: 
only supercritical (``productive") domains
evolve into lipid vesicles that are extracted from the membrane, while subcritical (``unproductive") domains
are rapidly dissolved~\cite{WCM+20,FPP+22}.
This perspective
fits within the broader framework of the far-from equilibrium formation
of biomolecular aggregates with specific functions, including 
lipid rafts~\cite{For05,FSH10b} and
specialized 
lipid-protein nanodomains~\cite{Des10,MD11,For12,BMD16,CCP+24}, such as 
cadherin and integrin clusters~\cite{QMM+13,BBG+24}.

Phase separation is emerging as one of the main ordering processes
in living cells~\cite{HWJ14,SB17,FPA+21}, and various mechanisms have been proposed as its drivers.
Among them, weakly polar electrostatic interactions between 
disordered regions of proteins~{\cite{HHR+17}},
active processes,
as in diffusion-limited phase separation, 
mass-conserved reaction-diffusion
systems and active 
emulsions~\cite{GCT+05,GKL+07,HBF18,BHF20,WZJL19,SDP+22,ZCT+15},
and segregating kinetic effects~\cite{WWF16}.	
Moreover, it has long been established that
protein inclusions in lipid membranes are subject 
to 
membrane-mediated interactions. These can originate
either
from ground state deformation of membrane shape, 
when protein inclusions 
are a source of
intrinsic curvature, 
or from membrane fluctuations, as the presence 
of embedded protein inclusions restricts membrane fluctuation modes, generating 
entropic
interactions~\cite{GBP93,GBP93b,W18}. 
We 
focus 
here
on 
the
latter class of interactions,
commonly known 
as Casimir-like forces. 
These
are non-additive, 
weak forces 
that are
mainly relevant
at short separations~\cite{DF99,LZM+11,YD12}.
At thermodynamic equilibrium, 
they are however sufficient to
induce 
a demixing transition 
in heterogeneous membranes~\cite{Wei02}.

It is known that proteins and lipids involved in the formation of sorting domains
increase local membrane rigidity by a factor 
of~10 to~30 
compared to the surrounding
membrane~{\cite{JPS+06,ZBM08,NCM+15}}. 
This suggests 
that 
entropic
forces may play 
a relevant
role
in the molecular sorting process. 
Here
we perform a thorough analysis of the problem and find
 that 
entropic forces
 significantly
enhance 
molecular sorting 
efficiency,
especially 
in the biologically 
relevant regime of weak 
direct
interactions.

\section{Phenomenological Theory}
\label{sec:phentheory}
Building on our previous work, we investigate the role of the lipid membrane as a
distiller of molecular species~{\cite{ZVS+21,FPP+22,PFA+23}}. In this
scenario, molecules are randomly inserted into the membrane, diffuse laterally,
and aggregate into sorting domains due to 
the action of
attractive forces. The sorting domains grow by
adsorbing molecules from the surrounding ``gas'' of freely diffusing molecules. Domains of size $R$
larger than a critical value $R_\mathrm{c}$ grow irreversibly
through the absorption of single molecules diffusing toward them~\cite{LS59,Sle09,FPP+22}. The growth
rate is determined by the net flux $\Phi$ of molecules toward a domain, which in
turn is proportional to the molecular density 
difference $\Delta n=n_L-n_R$ between distant regions and regions
adjacent to the domain boundaries~{\cite{ZVS+21}}. Domains that reach a characteristic
size $R_\mathrm{E}$ are ultimately removed from the membrane through the formation of
small, separate lipid vesicles~{\cite{ZVS+21}}.
It is worth observing here that 
vesicle formation is a complex process involving 
the concomitant action of a wide variety of genes,
as reviewed, for instance, in Ref.~\cite{KR18}. In our
approach, we abstract on molecular details and encode the mesoscopic effect
of vesicle extraction in the single parameter $R_\mathrm{E}$.

Of particular interest is the stationary out-of-equilibrium regime, where
molecular insertion and extraction processes are balanced. This balance can be described by the equation
\begin{eqnarray}
	\phi &=& N_\mathrm{d}\, \Phi,
	\label{eq:balance}
\end{eqnarray}
where $\phi$ is the flux density of molecules being inserted
into the membrane, $N_\mathrm{d}$ is the density of supercritical domains,
and $\Phi$ is the average flux of the molecules into a domain. In this regime, unlike in
the classical Lifshitz-Slezov scenario~{\cite{LS59,Sle09}}, the flux-driving jump $\Delta n$
in molecular density is kept finite by the continuous influx~$\phi$ of molecules into the membrane.

We have shown in Ref.~\cite{ZVS+21} that an optimal sorting regime is achieved
for an intermediate strength of 
  the
   attractive forces. When the tendency
to aggregate is too strong, a proliferation of slowly growing sorting
domains occurs, leading to molecular crowding and decreased
sorting efficiency~{\cite{ZVS+21,PFA+23}}. In the optimal sorting regime, there exists a
specific density $N_\mathrm{d}$ of sorting domains, resulting in minimal
average molecular density~{\cite{ZVS+21}}. For absorbing
domains, the average residence time $T$ of a molecule of linear size $a$ in the
membrane system is the sum of the average time $T_\mathrm{f}$ required for the
molecule to reach a sorting domain by free diffusion and be absorbed, and the
average time $T_\mathrm{d}$ spent inside the domain until the extraction
event. The two contributions can be estimated as~{\cite{ZVS+21}}
\begin{eqnarray*}
	T_\mathrm{f} \sim  \frac{1}{DN_\mathrm{d}}, \quad
	T_\mathrm{d} \sim \frac{(R_\mathrm{E} / a)^2}{\phi} N_\mathrm{d},
\end{eqnarray*}
where $D$ is the molecular diffusion coefficient.
The sum $T=T_\mathrm{f}+T_\mathrm{d}$ has a minimum for
\begin{eqnarray}
	N_{\mathrm{d}, \mathrm{opt}} & \sim & \frac{a}{R_\mathrm{E}}  \sqrt{\frac{\phi}{D}} .
	\label{eq:optnd}
\end{eqnarray}
The actual density $N_\mathrm{d}$ is a function of the microscopic properties of the system
that control the nucleation and growth of domains in the stationary state, but
irrespective of the combination of these microscopic quantities, the optimal residence time of molecules on the membrane
has the value determined by Eq.~(\ref{eq:optnd}).

To account for the role of membrane fluctuations in the molecular sorting process described above,
we recall that the equilibrium thermal fluctuations of an elastic membrane
are described by the Helfrich Hamiltonian,
\begin{eqnarray}
	\mathcal{H} & = & \int \mathrm{d}S  \left[ \frac{\kappa}{2}
	\left(\frac{1}{R_1}+\frac{1}{R_2}\right)^2 + \frac{\bar{\kappa}}{R_1R_2}
	\right],
	\label{eq:ham}
\end{eqnarray}
where the integral runs over the membrane surface, $\mathrm{d}S$ is the area element, $R_1,R_2$ are local principal curvature radii, and $\kappa,\bar{\kappa}$
are the bending rigidities associated with the mean and Gaussian curvatures,
respectively~{\cite{Can70,Hel73,LL86b}}.
As argued in Refs. \cite{HBD12,DKP+13,Nog20}, for biological membranes, $\overline{\kappa}$ is close to $-\kappa$. While our theory
remains valid for any relation between $\kappa$ and $\overline{\kappa}$, for
simplicity we will assume that
$\overline{\kappa}= - \kappa$ in the numerical computations presented in the following section.
In the presence of protein inclusions, the rigidity of the membrane becomes
spatially non-uniform.
Here, we assume that $\kappa (\mathbf{r}) = \kappa_0$ for the bulk membrane, and $\kappa (\mathbf{r}) = \kappa_1$ in the regions occupied by the molecules.
A~surface-tension contribution to the energy could
also be included, but it is assumed to be negligible and will not be considered~here.

We further assume that the diffusive dynamics of protein
inclusions is slower
than the fluctuational dynamics of the underlying membrane, i.e.,
$\tau_\mathrm{diff} \gg  \tau_\mathrm{rel}$,
with $\tau_\mathrm{diff}$ the characteristic
diffusion time and $\tau_\mathrm{rel}$ the characteristic membrane relaxation time.
This is motivated by the following estimates.
The characteristic time for lateral diffusion can be estimated as $\tau_\mathrm{diff}\sim{\lambda^2}/{D}$,
where~$\lambda$ is the characteristic scale of the problem.
Assuming that the viscosity $\eta$ of the cytosol is the primary
source of dissipation, the characteristic relaxation time
of the membrane dynamics is $\tau_\mathrm{rel} \sim {\eta \lambda^3}/{\kappa}$~\cite{BL75}.
Since the ratio $\tau_\mathrm{rel}/\tau_\mathrm{diff}$ increases as $\lambda$ grows,
one should check whether the inequality $\tau_\mathrm{diff} \gg  \tau_\mathrm{rel}$
holds for the largest characteristic scale, that is, for the size of the membrane.
Considering membranes with sizes $\lambda=100-500\,$nm, taking
the viscosity $\eta\sim 5\cdot10^{-3}\mathrm{Pa}\cdot\mathrm{s}$ and
the lateral diffusivity $D$ of proteins in the range $1 - 10$~$\mu$m$^2$/s  \cite{RHK09,WN13}, one finds
that the ratio $\tau_\mathrm{diff}/\tau_\mathrm{rel}$
spans the values $1-10^2$, suggesting that the dynamics of membrane fluctuations in living cells is
faster than lateral particle diffusion~\cite{BL75,NB07, DBC+02}.

\subsection{Interaction with a sorting domain}
Membrane fluctuations are known to induce 
effective interactions between inclusions within the membrane.
These interactions can be conveniently studied in the weak fluctuation regime,
where quantitative analyses can be
performed~{\cite{GBP93,GBP93b,PL96,BDF10,LZM+11,YD12}}.
It is of particular interest to investigate how these forces interplay with 
direct
forces
to facilitate the absorption of neighboring molecules by sorting domains.
In the
adiabatic approximation, 
justified by the timescale separation
$\tau_\mathrm{diff}\gg\tau_\mathrm{rel}$,
molecules included within the membrane
experience
effective
forces that 
can be computed by averaging over membrane fluctuations sampled from
the equilibrium distribution. 

Analytic expressions for membrane-mediated forces can be derived in various limit cases.
We are interested here in the interaction of a circular domain of size $R$
with a molecule of linear size $a$ situated at a distance~$x$ from it.
Approximating the domain boundary
in zeroth order as an infinite straight wall under the condition $R \gg x \gg a$,
the effective potential energy of the membrane-mediated
interactions is given by:
\begin{eqnarray}
	U (x) & = & - A \; k_\mathrm{B} T \hspace{0.17em} \frac{a^2}{x^2}
	\label{eq:near}
\end{eqnarray}
where $A$ is a dimensionless, increasing function of the relative rigidity
$\alpha =\kappa_1 / \kappa_0$ (see Appendix~\ref{sec:Interaction}). Eq.~(\ref{eq:near}) 
implies that $U \sim A\; k_\mathrm{B} T$ near the
surface of a domain. On the other hand, the interaction potential between two inclusions mediated
by the membrane fluctuations decays as $r^{- 4}$ for distances $r$ much larger than their sizes~\cite{GBP93}.
Notice that when considering a membrane surface tension $\sigma$, 
a new
relevant lengthscale, $\xi\sim\sqrt{\kappa/\sigma}$, 
emerges~\cite{Fou24,Leb22,BCF18}.
At~scales below $\xi$, surface tension 
has a weak influence on
membrane properties, 
whereas for scales above~$\xi$, it significantly modifies the long-range 
part
of 
the entropic 
interaction~\cite{BCF18, LZM+11,Leb22,Fou24}. 
As 
discussed in Refs.~\cite{LZM+11,YD12} and
in Appendix~\ref{sec:Interaction}, 
the entropic interaction
is 
mainly 
appreciable
at short separations. 
Therefore,
we expect
the effects of surface tension 
to be negligible in the present context.

\subsection{Sorting process}
The process of lateral diffusion of a molecule situated near a
circular sorting domain
can be described by the biased Brownian motion
\begin{eqnarray*}
	\dot{\mathbf{r}} & = & - \beta D \nabla U (\mathbf{r}) +\boldsymbol{\xi},
\end{eqnarray*}
where $\beta = (k_\mathrm{B} T)^{-1}$.
According to the fluctuation-dissipation theorem, the noise term $\boldsymbol{\xi}$ satisfies
\begin{eqnarray*}
	\langle \xi_i (t) \rangle & = & 0\\
	\langle \xi_i (t) \xi_j (t) \rangle & = & 2 \hspace{0.17em} D
	\hspace{0.17em} \delta_{ij}  \hspace{0.17em} \delta (t - t').
\end{eqnarray*}
It is worth observing here that in the limit of weak fluctuations, geometric effects caused
by the projection of the molecule's path can be neglected~{\cite{RS05,RLS10}}.
Moreover, deviations of the domain shape from circularity produce rapidly decaying higher
multipole contributions that may be neglected in the main approximation.

The time-dependent density profile of a population of such diffusing molecules around a domain obeys the following diffusion equation
\begin{eqnarray}
	\partial_t n (\mathbf{r}, t) & = & \nabla \cdot [D (\nabla + \beta \nabla U)
	n (\mathbf{r}, t)]  \label{eq:FP}
\end{eqnarray}
where $n$ is the two-dimensional molecular density.
To study the growth of the domain, one can consider
an isotropic, time-independent solution to Eq.~(\ref{eq:FP}).
The assumption of isotropy is justified by the circular
shape of the domain, while the approximate time independence
is supported by the slow nature of the diffusion process.
Consequently, $n$~and~$U$ depend only on the distance $r$ from the center of the domain.
The explicit expression for $n(r)$ is given~by:
\begin{eqnarray}
	n (r) & = &
	n(R) \exp\left[\beta U(R) -\beta U (r)\right]
	+
	\frac{\Phi}{2 \pi D}
	\int_R^r \frac{\mathrm{d} \rho}{\rho}
	\exp\left[\beta U (\rho) -\beta U (r)\right],
	\label{eq:FP2}
\end{eqnarray}
where $R$ is the radius of the domain and $n(R)$ is the molecular density near
the domain boundary. 
For realistic 
values 
$\alpha\sim 10-30$~{\cite{JPS+06,ZBM08,NCM+15}},
the potential $U$, induced by membrane fluctuations, is
at least
of the order of $k_\mathrm{B} T$ when $r\sim R$ and tends to zero as $r$ grows 
(see Appendix~\ref{sec:Interaction}).
The potential $U(r)$ 
rapidly 
approaches
zero when $r$ becomes much larger 
than $R$ (as $\sim (r/R)^{-4}$, 
see 
Eq.~(\ref{eq:inter})). 
This allows us 
to neglect $U(r)$ in Eq.~(\ref{eq:FP2}) when $r\gg R$. 
The leading 
contribution 
in $r/R$ 
can be extracted
by integrating by parts in the integral in Eq.~(\ref{eq:FP2}):
\begin{equation}\label{logcon}
{\cal J}=\int_{R}^{r}\frac{\mathrm{d}\rho}{\rho} \mathrm{e}^{\beta U(\rho)}=
\ln\frac{r}{R}+\delta\!\mathcal{J},
\end{equation}
where 
$\delta\!\mathcal{J}$ converges as $r\to \infty$:
\[
\delta\!\mathcal{J}\approx -\beta\int_{R}^{\infty}\mathrm{d}\rho 
\frac{\mathrm{d}U(\rho)}{\mathrm{d}\rho}\ln\left(\frac{\rho}{R}\right)
 \mathrm{e}^{\beta U(\rho)}.
\]
Since 
$|\delta\!\mathcal{J}|\le (\pi A)^{1/2} a/R \ll 1 $ for $a\ll R$,
it
can be neglected, 
leading to the relation
\begin{eqnarray}
	n(r)= n(R)\, \mathrm{e}^{\beta U(R)}
	+\frac{\Phi}{2 \pi D} \ln\frac{r}{\tilde R},
	\label{eq:FP3}
\end{eqnarray}
where
$\tilde R \sim R$. For the attractive potential induced by membrane fluctuations,
$U<0$, $\tilde R>R$ and $\tilde R-R\sim R$. 
The factor $\mathrm{e}^{\beta U(R)}$ in Eq.~(\ref{eq:FP3}) is of order unity.

The density of molecules near the domain boundary is determined by the dynamic equilibrium 
of association and dissociation processes and,
using the Gibbs-Thomson relation~\cite{LL80,KRN10}, 
can be expressed as
\begin{eqnarray}
	n(R)=n_0 (1+R_\star/R),
	\label{eq:FP4}
\end{eqnarray}
where $n_0$ is the equilibrium density near a straight boundary, 
and the $R$-dependent correction accounts for the
effect of linear tension. This correction is directly related to the curvature of the domain boundary. The length
$R_\star$ in Eq.~(\ref{eq:FP4}) can be estimated to be of the order of a few 
molecular radii. Expression~(\ref{eq:FP4})
allows to determine the critical radius $R_\mathrm{c}$: by definition, a domain with radius $R_\mathrm{c}$ remains static, since the flux
$\Phi$ for such a domain is zero. Substituting $\Phi=0$ and $n(r)=n_L$ (where $n_L$ is the concentration of the
molecules far from the domains) into Eq.~(\ref{eq:FP3}) yields:
\begin{eqnarray}
	\frac{R_\star}{R_\mathrm{c}}
	&=&\frac{n_L}{n_0}\, \mathe^{-\beta U} -1.
	\label{eq:FP5}
\end{eqnarray}
Since $\exp(-\beta U)>1$ for the attractive potential, we conclude from Eq.~(\ref{eq:FP5}) that membrane-induced attraction reduces the
critical radius. For domains larger than $R_\mathrm{c}$, the correction related to the linear tension can be neglected, resulting
in $n(R)\to n_0$. Consequently, we find from Eq.~(\ref{eq:FP3}):
\begin{eqnarray}
	n_L- n_0\, \mathrm{e}^{\beta U}
	=\frac{\Phi}{2 \pi D} \ln\frac{L}{\tilde R},
	\label{eq:FP6}
\end{eqnarray}
where $L$ is a distance of the order of the separation between the domains. Since $\exp(\beta U)<1$ for the attractive
potential, we conclude from Eq.~(\ref{eq:FP6}) that membrane-mediated attraction enhances the effectiveness of the clustering process,
resulting in an increased flux $\Phi$.

The above relations show how forces mediated by membrane fluctuations affect the sorting process.
Let us examine the effect of increasing membrane-mediated attraction
(which can be directly adjusted in numerical simulations by
varying the relative rigidity $\alpha=\kappa_1/\kappa_0$).
As  membrane-mediated attraction increases, the critical radius
$R_\mathrm{c}$ of the domains decreases, leading to a higher
rate of production of germs of new sorting domains
and, consequently, an increased overall density~$N_\mathrm{d}$ of sorting domains~\cite{Sle09}.
However, according to the balance relation (\ref{eq:balance}), this should concomitantly
result in a lower $\Phi$ and, in accordance with
Eq.~(\ref{eq:FP6}), a lower $n_L$, which in  turn reduces
the rate of new domain generation. Between these two opposing effects, the first
is expected to dominate due to the high sensitivity of the process of germ generation to
the critical radius $R_\mathrm{c}$~\cite{Sle09}. 
The following physical picture thus emerges.
The efficiency of the sorting process is controlled by the rate of nucleation 
of new sorting domains. 
According to 
classical nucleation theory 
the
rate
of generation of new sorting domains depends exponentially on
$R_\mathrm{c}$~\cite{Sle09}. 
Eq.~(\ref{eq:FP5})
implies
that even entropic interactions  
$\beta U\sim 1$ 
significantly affect 
$R_\mathrm{c}$.
At short distances, 
the entropic force 
acts as a facilitator
of nucleation by 
biasing molecular diffusion toward sorting domains and stabilizing them.
While at equilibrium a sharp demixing transition is observed above a critical
value of rigidity~\cite{Wei02}, in the statistical steady state
of interest here
we
expect to 
observe 
a 
smooth increase of the rate of nucleation of sorting domains
with increasing rigidity of the inclusions.
A~significant effect is expected
in particular
in the realistic 
range $\alpha\sim 10-30$~{\cite{JPS+06,ZBM08,NCM+15}}, where
$\beta U\sim 1$ in the proximity of the domains.

It is worth 
observing here
that 
in the 
statistical steady
state, the density $N_\mathrm{d}$ of
sorting domains is self-consistently determined through the stationarity
condition $\mathd N_\mathrm{d} / \mathd t = \phi / N_\mathrm{E}$, 
where $N_\mathrm{E}$ is the average number
of molecules removed during an extraction event,
since the rate of formation of new domains
$\mathd N_\mathrm{d} / \mathd t$
is in average equal to
the rate of extraction events.~\cite{ZVS+21,FPP+22}.
Starting from the regime of
weak 
direct
interactions, the optimal density of sorting domains $N_{d,\tmop{opt}}$ determined by Eq.~(\ref{eq:optnd})
can be reached either by increasing the 
direct
interaction strength, or by reducing the critical
radius by means of increased molecular rigidities $\kappa_1$. Conversely, to
increased molecular rigidities should correspond lower values of the optimal 
direct
interaction strength.

\section{Numerical results}
\begin{figure}[!b]
	\centering
	\includegraphics[width=0.41\textwidth]{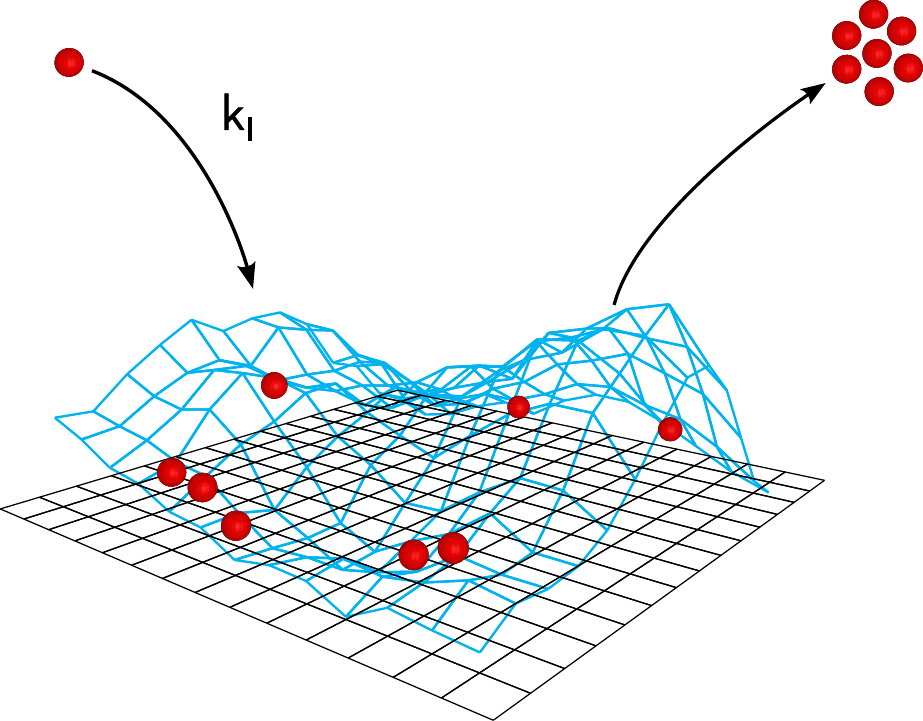}
	\caption{
		Schematic representation of the discrete model of molecular sorting on a
		fluctuating membrane. The membrane (in blue) is
		described by its height relative to a reference plane (in black).
		Rigid molecules are inserted into vacant
		sites at a rate $k_\mathrm{I}$, and
		connected domains containing more molecules than the
		threshold size $N_\mathrm{E}$ are extracted.
		The amplitude of membrane fluctuations is here amplified for
		clarity. }
	\label{fig:schematic}
\end{figure}
To validate our theoretical predictions, we
implemented a numerical scheme that generalizes the lattice-gas model of
molecular sorting introduced in Ref.~\cite{ZVS+21}. This scheme
shares several features with the approach used in Ref.~\cite{Wei02} to investigate
the phase separation of rigid inclusions in fluid membranes close to
thermodynamic equilibrium, although we are studying here an out-of-equilibrium state. We consider a fluctuating membrane described by a
discretized version of Helfrich Hamiltonian, on which inserted molecules
laterally diffuse and aggregate. The system is driven out of equilibrium by an
incoming flux of molecules, which are randomly attached at empty membrane sites with a
rate $\phi$ per unit area, and is maintained in a statistical stationary state by the instantaneous
removal of connected molecular domains that reach
the threshold number of molecules~$N_\mathrm{E}$.
Consistently with our theoretical approach, simulations are performed
in the adiabatic regime.

In our numerical scheme, the membrane configuration is described by the
height $u_{i}$ of its points relative to a reference plane, which
is discretized into a square lattice of $L \times L$ sites,
see Fig.~\ref{fig:schematic}.
To avoid boundary effects, periodic boundaries conditions are applied.
Each site of the lattice can be occupied by at most one molecule.
An occupation number $n_{i} \in \{0,1\}$ is associated to each site~$i$. Sites
with $n_{i} = 0$ have the
bending rigidity $\kappa_0$, while sites with
$n_{i} = 1$ have the rigidity $\kappa_1$. The
corresponding Gaussian rigidities are assumed to be equal to
$-\kappa_0$ and $-\kappa_1$, respectively.
To account for the 
direct
attractive force
between membrane inclusions we add to the discretized Helfrich energy of the membrane
the nearest-neighbor interaction energy
\begin{eqnarray}
	H_\mathrm{incl}
	&=&
	-\frac{W}{2}
	\sum_{\left< i,j\right>}  n_{i} n_{j}
	\label{eq:discreteH}
\end{eqnarray}
Membrane configurations are sampled using a Monte Carlo algorithm. After each Monte Carlo sweep (MCS), steps involving molecule insertion, diffusion, and the
extraction of domains of size $\ge N_\mathrm{E}$ are
performed. One MCS is taken as the time unit.
The rate of molecule insertion per empty site is denoted by $k_\mathrm{I}$.
The diffusion rate~$k_\mathrm{D}$ of free molecules is
measured as the ratio of accepted diffusive jumps during one MCS
(see Appendix~\ref{sec:MonteCarlo} for additional details).
Simulations are performed with the realistic parameter values
$\kappa_0 = 10\, k_\mathrm{B} T$, 
$N_{\mathrm{E}} = 25$~\cite{P10,Rawicz2000,ZVS+21,FPP+22,PFA+23},
while $k_I$ and $k_D$ are kept much smaller than $1$
in inverse MCS units, to ensure proper sampling of membrane
configurations within the adiabatic regime.
To match simulation parameters with real-world units, we consider
that each square plaquette in the lattice corresponds to a patch of lipids of 
area $h^2$, with $h\approx 10\,$nm, the order of magnitude of the lateral size of 
typical protein inclusions, and also of the shortest fluctuational wavelengths 
for membrane bending deformations~\cite{Wei02,SNF16}. For molecular
diffusivities $D\approx 1\;\mu\mathrm{m}^2/\mathrm{s}$, the typical time between 
consecutive diffusive jumps of a free inclusion on the lattice is
$k_D^{-1}=h^2/D\approx 10^{-4}\;\mathrm{s}$. 

The average density $\bar{\rho}$ of molecules in the
stationary state satisfies the relation $\bar{\rho} = \phi\,
T$, where $T$ is the average time a particle spends on the membrane before being extracted, and
$\phi = k_\mathrm{I}  (1 - \rho)$ is the flux
of incoming particles per site, if lengths are measured in units
of the lattice spacing~\cite{ZDG19,ZVS+21}.
Therefore, in the statistically stationary state established at fixed $\phi$,
the average density $\bar{\rho}$ is a measure of the efficiency of the sorting process~{\cite{ZVS+21}.}

\begin{figure}[!b]
	\centering
	\includegraphics[width=0.47\textwidth]{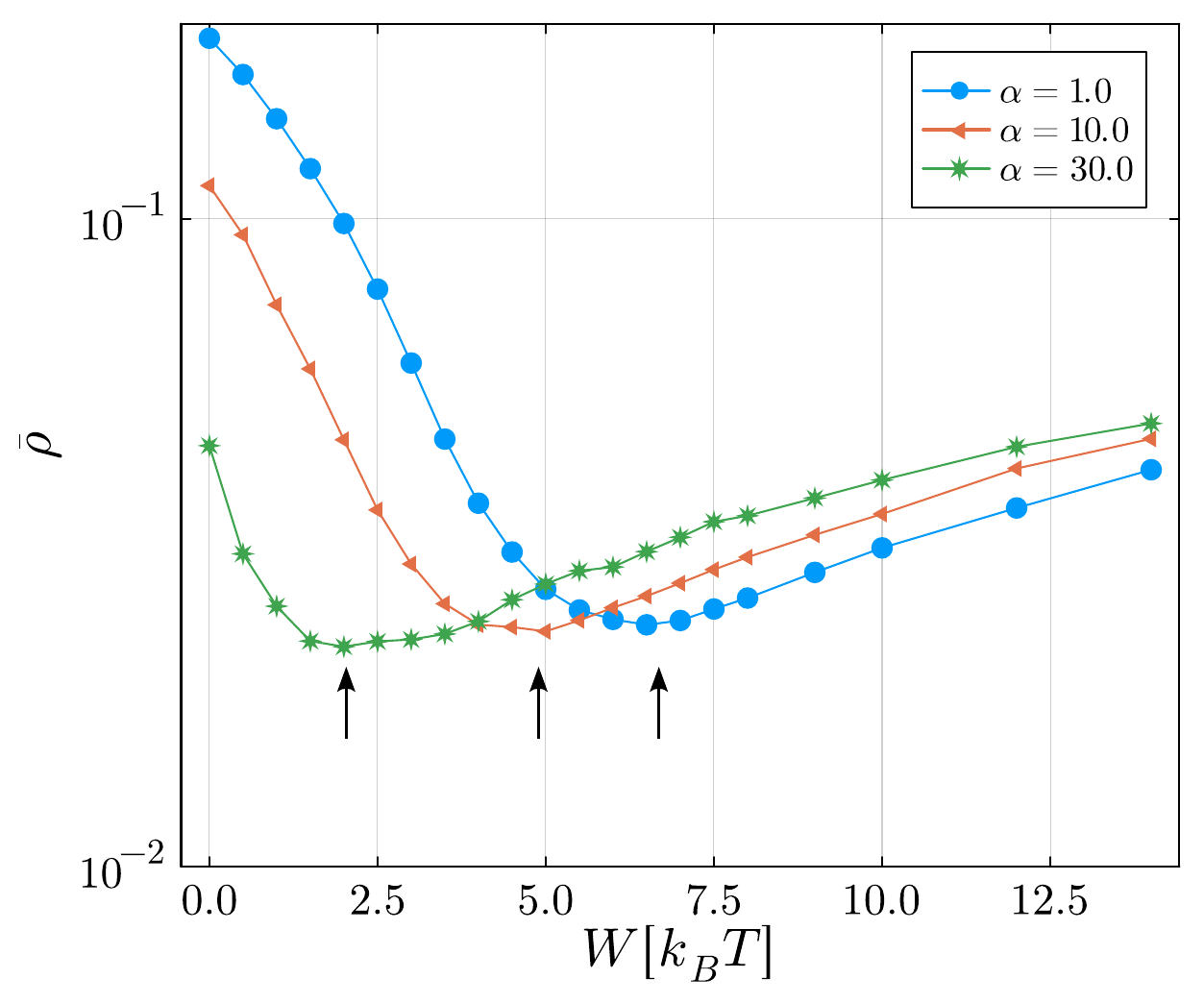}
	\caption{Average density $\bar{\rho}$
		in the stationary state as a function of the 
direct
      interaction strength $W$.  The different curves correspond
		to different values of
		$\alpha=\kappa_1/\kappa_0$. 
		The optimal sorting region depends on
		both the 
direct
      interaction and the rigidity of the
		biomolecules involved.
      For
      larger values of the relative rigidity $\alpha$, 
      the density curve and 
      the optimal interaction
      strength $W_\mathrm{opt}$ 
      shift 
      toward lower values of~$W$.
Simulations were performed with $\phi/k_\mathrm{D}=10^{-5}$,
$\kappa_0=-\bar{\kappa}_0=10\,k_\mathrm{B}T$, 
$\kappa_1=-\bar{\kappa}_1$,
$L=100$, $N_\mathrm{E}=25$.
For $h=10\,$nm and $D=1\,\mu\mathrm{m}^2/\mathrm{s}$, one 
has
$k_D^{-1}=10^{-4}\,$s. 
      }
	\label{fig:density-W}
\end{figure}
\begin{figure*}[!b]
	\centering
	\includegraphics[width=1.0\textwidth]{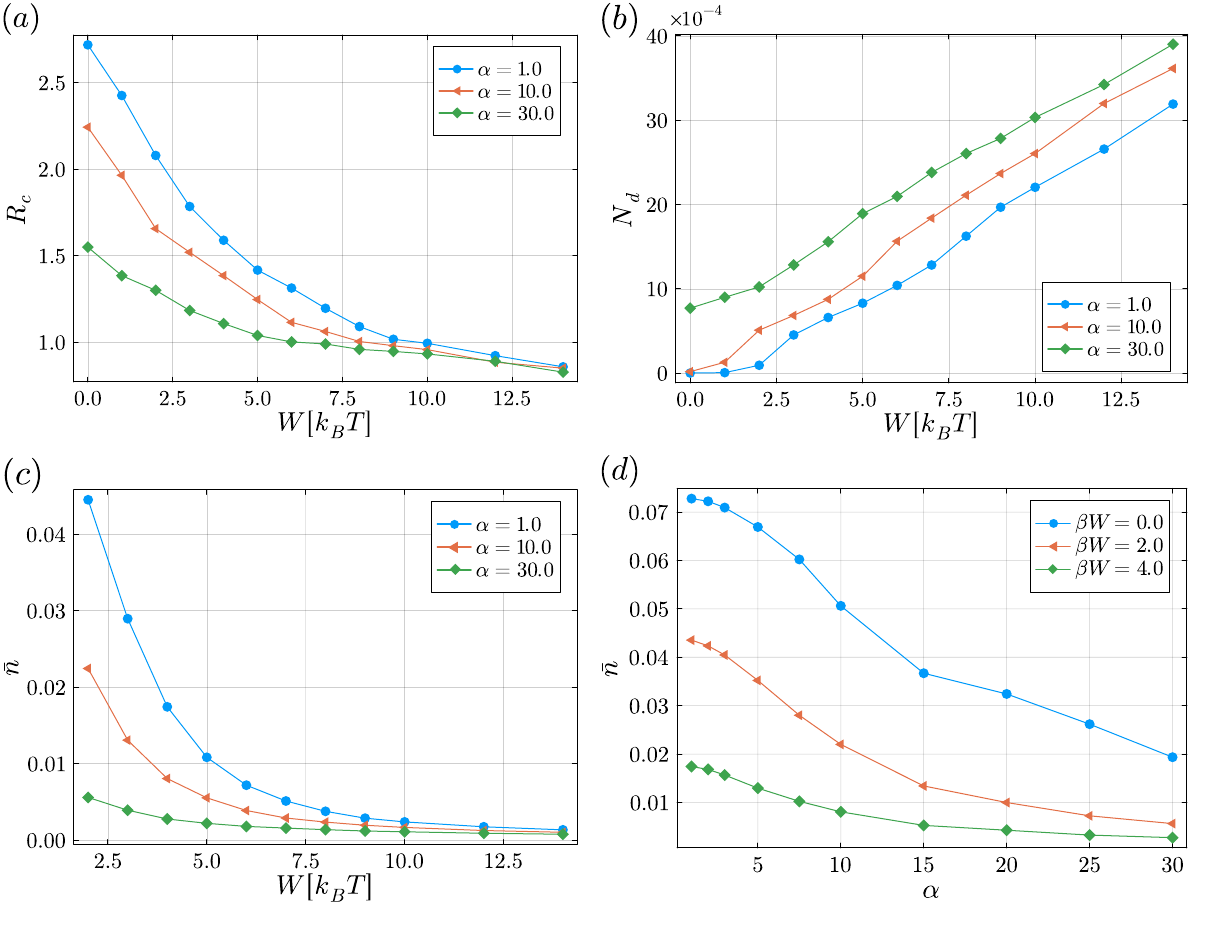}
	\caption{Characterization of the sorting process
		in the statistically
		steady state in terms of three
		key observables, measured from numerical simulations as functions of the 
		direct
		interaction
		strength $W$, for varying relative rigidities $\alpha=\kappa_1/\kappa_0$:
		{(a)} the critical radius $R_\mathrm{c}$ (estimated
		using the method described in Ref.~\cite{FPP+22});
		{(b)} the number density $N_\mathrm{d}$ of supercritical
		domains;
		{(c)} the average density of isolated molecules $\bar{n}$;
		and {(d)} the average density of isolated molecules $\bar{n}$ as a function of $\alpha$ for three
		different values of $W$.
		Due to the logarithmic profile of the molecular density around
		sorting domains, the average density $\bar{n}$ is
		close to
		$n_L$.
		Same parameters as in Fig.~\ref{fig:density-W}.
	}
	\label{fig:panel}
\end{figure*}

We investigated the behavior of the density $\bar{\rho}$ as a function of the
direct
interaction $W$ and molecular rigidity $\kappa_1$. In
Fig.~\ref{fig:density-W}, the resulting stationary densities are plotted as
functions of the 
direct
interaction strength $W$ for the fixed dimensionless flux
$\phi / k_\mathrm{D} = 10^{- 5}$ (see Appendix~\ref{sec:MonteCarlo}),
with varying $\alpha =\kappa_1 / \kappa_0$.

These numerical results confirm the theoretical prediction that
membrane-mediated interactions strongly influence the molecular sorting process,
and that the optimal 
direct
interaction
strength $W_\mathrm{opt}$ 
decreases as the intensity of membrane-mediated
interactions increases 
(Fig.~\ref{fig:density-W}), 
thus enhancing sorting efficiency
in the biologically relevant regime of weak 
direct
interactions.
Since the entropic force mainly acts at short separations,
it~renormalizes the value of the direct interaction,
resulting in an effective short-range interaction strength~$W_\mathrm{eff}$,
as 
evidenced by
the consistent shift of the density curves toward lower values of~$W$
in~Fig.~\ref{fig:density-W}.
However, 
it is important to note 
that the entropic component of $W_\mathrm{eff}$ has 
a 
distinct
origin and parametric dependence 
compared to
the direct interaction part, 
as it is 
governed
by the relative rigidity $\alpha$.

To further validate the present theoretical scenario, we measured the critical size $R_\mathrm{c}$,
the number density of sorting domains $N_\mathrm{d}$, and the
average density of isolated molecules $\bar{n}$ (which is
approximately the same as~$n_L$) for varying values of~$W$
and~$\alpha$ (Fig.~\ref{fig:panel}). Consistent with the theoretical predictions,
the critical size $R_\mathrm{c}$ decreases monotonically with both
increasing $W$ and~$\alpha$ (Fig.~\ref{fig:panel}a),
resulting in a higher sorting domain density $N_\mathrm{d}$ (Fig.~\ref{fig:panel}b).
This confirms that, in the presence of membrane-mediated interactions,
the optimal sorting-domain density~$N_{\mathrm{d},\mathrm{opt}}$
is achieved at lower 
direct
interaction strengths~$W$. As predicted,
the increase in sorting-domain density is reflected
in a corresponding decrease in the average density of isolated molecules $\bar{n}$ (Fig.~\ref{fig:panel}c).
As expected, the effect of 
entropic
forces on the sorting process exhibits a smooth dependence on the 
relative rigidity $\alpha$, and it
becomes 
particularly significant 
for $\alpha \gtrsim 10$, where $\beta U\sim 1$ (Fig.~\ref{fig:panel}d).
Furthermore, 
we
tested
the effect of 
including
a surface term in our simulations
and observed
no significant change 
up to the
realistic value 
$\sigma = 10^{-5}\mathrm{J}\,\mathrm{m}^{-2}$~\cite{SNF16}, 
in line
with 
theoretical arguments.

\section{Conclusions}
The lipid membranes of endosomes,
the Golgi apparatus, the endoplasmic reticulum, and the plasma membrane
play a fundamental role in sorting and distilling vital molecular factors,
acting as a natural realization of Szilard's model of classical nucleation theory~{\cite{Sle09}}.
These delicate structures are inherently subject to thermally induced fluctuations.
Previous studies have shown that such fluctuations significantly contribute to the phase separation of rigid membrane inclusions
close to thermodynamic equilibrium~\cite{Wei02}.
Our analysis extends these findings to the out-of-equilibrium scenario of molecular sorting, demonstrating that
membrane-mediated interactions can strongly enhance the molecular distillation of
rigid inclusions, particularly, in the biologically
relevant regimes where 
direct
intermolecular attractive forces are relatively weak.
Our analysis suggests that thanks to membrane-mediated interactions, rigid biomolecules
can be sorted with  high efficiency, despite their low-affinity interactions.
Notably, this effect, potentially crucial for biological systems,
is observed in our numerical simulations well below the threshold
where phase separation occurs close to equilibrium~\cite{Wei02}.
This suggests an important distinction between classical quasi-equilibrium phase 
separation processes and the
role phase separation plays in out-of-equilibrium biological systems.
Note that from the 
point of view 
of 
macroscopic kinetics, 
the
entropic forces 
can be described by a
short-range
interaction 
of the same type 
as in
Eq.~(\ref{eq:discreteH}),
with 
an
effective parameter $W_\mathrm{eff}$, 
corresponding to attraction of particles toward the domains. 
The value of $W_\mathrm{eff}$ 
is determined by the 
relative rigidity
of the domain and has 
order 
$k_\mathrm{B}T$.
As a result, at low and moderate values of $W$, 
entropic 
forces 
provide a
universal 
physical
mechanism 
for 
the aggregation
of 
molecules 
within
the membrane,
independent of microscopic interaction details.
We believe that this 
should be 
taken into account 
in the design and interpretation of biological experiments.

Molecular inclusions interact with the surrounding membrane
due to both their rigidity and, possibly, non-zero intrinsic curvature~\cite{GBP93,PUWS09}.
In this study, we have focused on the impact of rigidity on the molecular sorting process. In future work, we plan to
investigate the complex interplay between rigidity and intrinsic curvature.

Here,
we did not account for the role of the cytoskeleton, which could influence the membrane fluctuation spectrum. At the
mesoscopic scale, a more accurate description can be achieved by incorporating additional terms into the membrane
Hamiltonian (Eq. \ref{eq:ham}) to account for interactions with the cytoskeletal network. Previous research has explored
this aspect by introducing local membrane pinning \cite{LB05,CYK15} or by considering membrane confinement \cite{GZS03}.
These studies found that at short wavelengths, the fluctuation spectrum of a free membrane is retrieved.
This suggests that 
the effect of the cytoskeleton on 
entropic
forces, which 
in the present context mainly act at short separations,
may be 
weak.
This point will be the subject of future investigations. 

Our findings suggest that a key parameter 
governing
molecular sorting efficiency is the relative rigidity
of the membrane and supermolecular domains,
which 
affects
the critical radius for
the nucleation of nascent sorting domains. 
The statistical distribution of domain sizes 
provides
an accessible
signature of many self-organized aggregation processes~\cite{QMM+13,BMD16}. 
The domain size distribution 
predicted by our model~\cite{ZVS+21}
aligns well with
experimental data 
for
endocytic sorting 
and can be used to infer the critical size~\cite{FPP+22}.
This suggests a practical way 
to experimentally investigate the 
physical picture of molecular sorting proposed in this work.

\section*{Acknowledgements}
	AG would like to thank Guido Serini 
	for many fruitful discussions.
	

\paragraph{Funding information}
Numerical calculations have been made possible through a CINECA-INFN agreement 
providing access to computational resources at CINECA.


\begin{appendix}
	\numberwithin{equation}{section}
	\section{Interaction of a molecule with a domain}
	\label{sec:Interaction}
	
	In this section, we analyze the 
   Casimir 
   interaction
	between a circular domain of radius~$R$ and a single molecule of  radius $a\ll R$,
	positioned at a distance $x\gg a$ from it.
	We will calculate the interaction potential between the molecule and the domain.
	
	In the absence of overhangs, the membrane can be parameterized in the Monge gauge~\cite{NPW04-book},
	where each point on the membrane is defined
	by its displacement $u (\mathbf{r}) = u (x, y)$ in the direction perpendicular to
	a reference plane~$\mathcal{S}$. To second order in $u$, the Helfrich Hamiltonian,
	which provides the elastic energy of the deformed membrane, reads
	\begin{align}
		\mathcal{H} & =  \int_{\mathcal{S}} 
		\mathrm{d}x\,\mathrm{d}y\, 
		\left\{ \frac{\kappa}{2}
		(\nabla^2 u)^2 + \bar{\kappa} [\partial_x^2 u\, \partial_y^2 u -
		(\partial_x\partial_y u)^2]
		\right\},
		\label{eq:ham_quadratic}
	\end{align}
	Here $\kappa$ and $\bar\kappa$ are bending and Gaussian rigidities, determined by an internal 
	structure of the membrane.
	A surface-tension contribution to the energy could also be included, but it is
	assumed to be negligible and will not be 
	taken into account.
	
	Here we consider the interaction of a single molecule with a circular domain of molecules inserted into the membrane.
	When the molecule is positioned at the point $\mathbf{r}=(x,y)$, the interaction potential
	of the molecule with the domain is
	\begin{eqnarray}
		U&=&\phantom{+}
		B
		(\partial_x^2\partial_{x'}^2\mathcal{G}
		|_{x=x',y=y'}
		+2\partial_x^2\partial_{y'}^2\mathcal{G}
		|_{x=x',y=y'}
		+\partial_y^2\partial_{y'}^2\mathcal{G}
		|_{x=x',y=y'}
		)
		\nonumber
		\\ &&
		+D
		(\partial_x^2\partial_{y'}^2 \mathcal{G}
		|_{x=x',y=y'}
		-\partial_x\partial_y\partial_{x'}\partial_{y'} \mathcal{G}
		|_{x=x',y=y'})
		\label{eq:potential}
	\end{eqnarray}
	where ${\mathcal G}(\mathbf{r} ,\mathbf{r}')$ is the contribution to the pair
	correlation function $\langle u(\mathbf{r}) u(\mathbf{r}')\rangle$
	from the membrane displacement induced by the domain.
	The factors $B,D$ in Eq.~(\ref{eq:potential}) are introduced via
	the phenomenological coupling energy of the molecule with the membrane,
	when the former is treated as a point-like object:
	\begin{equation}
		\delta\mathcal{H}=B(\nabla^2 u)^2
		+D[\partial_x^2 u
		\partial_y^2 u
		-(\partial_x\partial_y u
		)^2]
	\end{equation}
	where the derivatives are evaluated at the position of the molecule.
	This expression is valid for fluctuations of $u$ on scales much larger than $a$.
	The factors $B$ and $D$ are functions of
	the rigidity and size of the molecule. 
	We will make use of the fact that 
	their expression
	for a disc of radius $a$ and
	rigidity 
   $\kappa=\kappa_2,\; \bar\kappa=-\kappa_2$, inserted in a
	membrane of rigidity $\kappa=\kappa_0,\; \bar\kappa=-\kappa_0$ is~\cite{LZM+11,Leb22}:
	\begin{eqnarray}
		B
		&=&\pi a^2 \kappa_0(
      \kappa_2-\kappa_0)\left(\frac{1}{(
      \kappa_2+\kappa_0)}
		+\frac{1}{
      \kappa_2+3\kappa_0} \right)
		\nonumber  \\
		D
		&=&-\pi a^2 \frac{4(
      \kappa_2-\kappa_0)\kappa_0}{
      \kappa_2+3\kappa_0}.
		\label{eq:factorsBD}
	\end{eqnarray}
	
	If the separation between the molecule and the domain boundary is much smaller
	than the domain size $R$, the boundary can be approximated as a straight line.
	Therefore, we assume that the domain occupies the half-plane $x<0$.
	We also consider that the domain and the bulk membrane
	have different bending and Gaussian rigidities,
	$\kappa_1,\bar\kappa_1$ and $\kappa_0,\bar\kappa_0$. respectively.
	The Hamiltonian of the system is  then given by
	\begin{align}
		\mathcal{H}=\int_{\mathcal{D}_1} dx\, dy\, 
		\left\{ \frac{\kappa_1}{2} (\nabla^2 u)^2 + \bar{\kappa}_1 [\partial_x^2 u\partial_y^2 u -(\partial_x\partial_y u)^2]\right\}
		\nonumber \\
		+\int_{\mathcal{D}_2} dx\, dy\, 
		\left\{ \frac{\kappa_0}{2} (\nabla^2 u)^2 + \bar{\kappa}_0 [\partial_x^2 u\partial_y^2 u -(\partial_x\partial_y u)^2]\right\}
		\label{eq:hamhalf}
	\end{align}
	where $\mathcal{D}_1$ is the left half-plane ($x<0$) and $\mathcal{D}_2$ is the right half-plane ($x>0$).
	
	Using linear response theory, we can derive an equation for the
	pair correlation function $G=\langle u(\mathbf{r}) u(\mathbf{r}')\rangle$, entering~Eq.~(\ref{eq:potential}).
	It is important to note here that, due to the system's homogeneity in the $y$ direction
	and its invariance under reflection $y\to -y$, $G$ is a function of $|y-y'|$.
	The resulting equations read
	\begin{eqnarray}
		\nabla^4 G & = & \frac{k_\mathrm{B} T}{\kappa_1} 
		\delta(x-x')
		\delta(y-y') 
		\qquad x<0
		\nonumber \\
		\nabla^4 G & = &\frac{k_\mathrm{B} T}{\kappa_0} 
		\delta(x-x')
		\delta(y-y') 
		\qquad x>0
		\label{eq:eq_green}
	\end{eqnarray}
	with boundary conditions
	\begin{eqnarray}
		\partial_x(\kappa_1 \nabla^2 - \bar{\kappa}_1 
		\partial_y^2)
		G|_{x=0^-} 
		& = & \partial_x(\kappa_0 \nabla^2 - \bar{\kappa}_0 
		\partial_y^2)
		G|_{x=0^+}
		\nonumber \\
		(\kappa_1 \nabla^2 + \bar{\kappa}_1 
		\partial_y^2) 
		G|_{x=0^-} 
		& = &(\kappa_0 \nabla^2 + \bar{\kappa}_0 
		\partial_y^2) 
		G|_{x=0^+}
		\label{eq:cond_green}
	\end{eqnarray}
	Observe that, due to the inhomogeneity of the Gaussian rigidity, the topological term
	involving Gaussian curvature in the Hamiltonian cannot be neglected. This term
	contributes to the boundary conditions (\ref{eq:cond_green}) for the correlation function.
	
	Due to translation invariance along the $y$ direction,
	it is convenient to make use of the Fourier transform 
	\begin{equation}
		\hat G(x,x',q)=\int_{-\infty}^{+\infty} 
		\mathrm{d}y\,
		\exp[i q 
		(y-y')]
		G(x,x',
		y-y'),
		\nonumber
	\end{equation}   
	which is an even function of $q$. The solutions to
	Eqs.~(\ref{eq:eq_green}) and~(\ref{eq:cond_green}) for $q>0$ are
	\begin{equation}
		\hat{G}(x,x',q)=(A_0+A_1 x)\,\mathrm{e}^{qx}+\frac{k_\mathrm{B}T}{4 q^3
			\kappa_1
		}(1+q|x-x'|)\,\mathrm{e}^{-q|x-x'|}
		\nonumber
	\end{equation}
	for $x<0$, and
	\begin{equation}
		\hat{G}(x,x',q)=(B_0+B_1 x)\,\mathrm{e}^{-qx}+\frac{k_\mathrm{B}T}{4 q^3
			\kappa_0
		}(1+q|x-x'|)\,\mathrm{e}^{-q|x-x'|}
		\nonumber
	\end{equation}
	for $x>0$. The factors $A_0,A_1,B_0,B_1$
	must be determined from the continuity of $\hat G$ and its derivative $\partial_x \hat G$ at $x=0$
	and from the boundary conditions (\ref{eq:cond_green}), where $\partial_r^2\to -q^2$, $\nabla^2\to \partial_x^2 -q^2$. Assuming $\bar{\kappa}_0=-\kappa_0$ and $\bar{\kappa}_1=-\kappa_1$, the correlation function for $x,x'>0$ is
	\begin{eqnarray}
			\hat{G}(x,x',q) &=& \frac{k_\mathrm{B}T}{4 q^3 \kappa_0}\Biggl[ (1+q|x-x'|)
         \,\mathrm{e}^{-q|x-x'|}\nonumber\\
			&\phantom{=}&
			- \frac{\mathrm{e}^{-q(x+x')}(\kappa_1-\kappa_0)
         ((3\kappa_1+\kappa_0)(x+x'+2qxx')\,q+3\kappa_1+5\kappa_0)}{(3\kappa_1+\kappa_0)(\kappa_1+3\kappa_0)}  \Biggr].
			\label{eq:paircorr}
	\end{eqnarray}
	The second term in the square brackets determines
	the contribution $\mathcal G$ to the correlation function induced by the domain.
	
\begin{figure}[!b]
		\centering
		\includegraphics[width=0.55\textwidth]{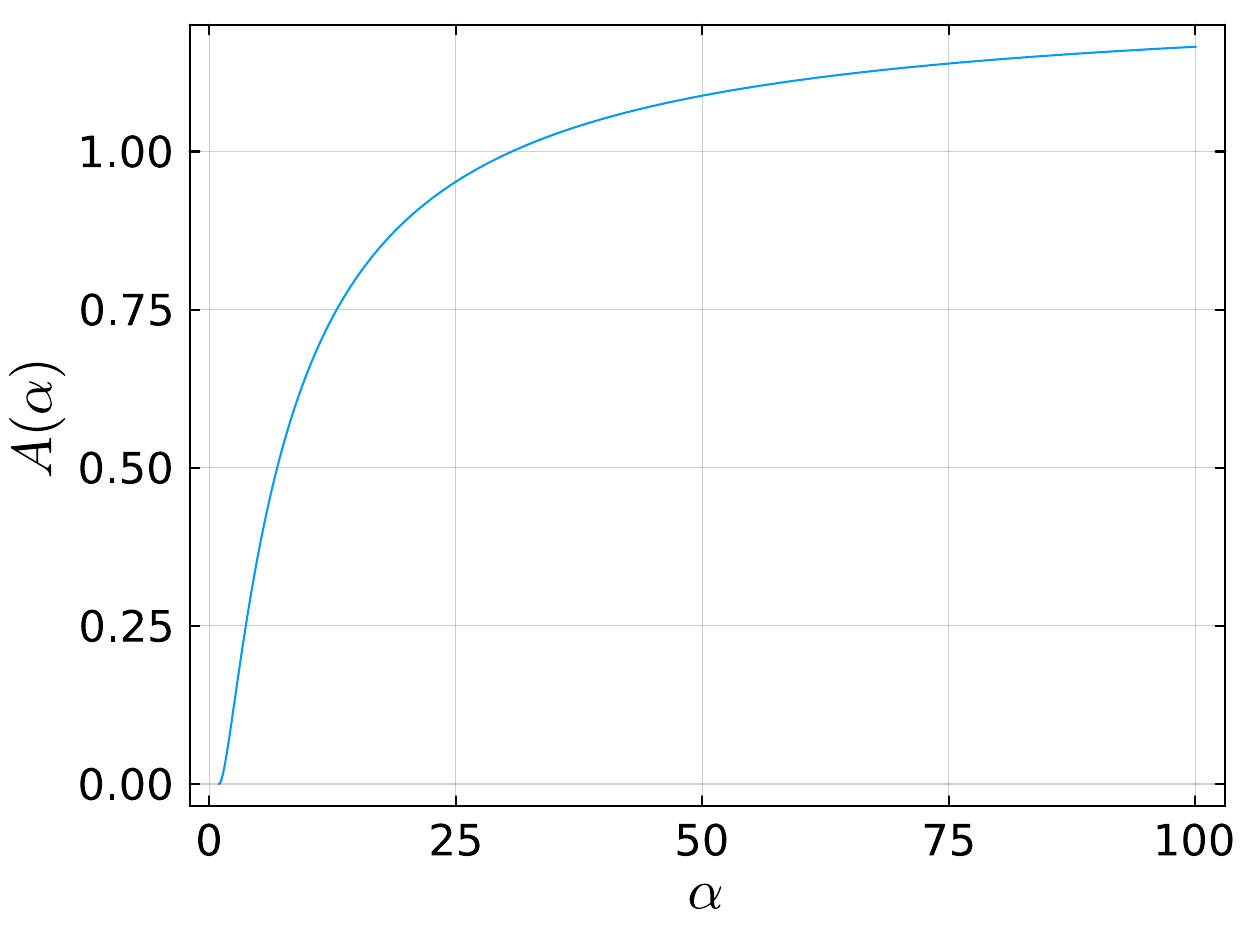}
		\caption{
        Dependence of the prefactor $A$ 
from
      Eq.~\ref{eq:factorA} on the relative rigidity~$\alpha$.
      }
		\label{fig:Aplot}
	\end{figure}

	In accordance with Eqs. (\ref{eq:potential},\ref{eq:factorsBD},\ref{eq:paircorr})
	the interaction energy of the molecule with the domain is
	\begin{equation}
		U(x)=  -k_{B}T\frac{\left(\kappa_{1}-\kappa_{0}\right)}{4(\kappa_{1}+3\kappa_{0})}\left(\frac{\kappa_{2}-\kappa_{0}}{\kappa_{2}+3\kappa_{0}}\right)\left[\frac{15\kappa_{2}\kappa_{1}+13\kappa_{2}\kappa_{0}+21\kappa_{0}\kappa_{1}+15\kappa_{0}^{2}}{(\kappa_{2}+\kappa_{0})(\kappa_{0}+3\kappa_{1})}\right]\frac{a^{2}}{x^2}
		\label{eq:general}
	\end{equation}
	When the molecule and the domain have the same rigidity ($\kappa_2=\kappa_1$), we obtain
	\begin{equation}
		\label{eq:near_app}
		U(x)=- A\, k_\mathrm{B}T\, \frac{a^2}{x^2}
	\end{equation}
	where, letting $\alpha=\kappa_1/\kappa_0$,
	\begin{equation}
		A  =  \frac{(\alpha - 1)^2  (3 \alpha + 5) (5 \alpha + 3)}{4 (\alpha + 1)
			(\alpha + 3)^2  (3 \alpha + 1)}
		\label{eq:factorA}
	\end{equation}
   which
   is a monotonically increasing function
   for $\alpha>1$,
   taking on values of order~1 for $\alpha\gtrsim 10$ 
   (Fig.~\ref{fig:Aplot}).
	
	\paragraph{Interaction at large distances}
	At large separations between the molecule and the domain, the size $R$ of the domain becomes a relevant scale, and its boundary can no longer be treated
	as an infinite wall. In this case, the interaction can be evaluated as \cite{Leb22}
	\begin{eqnarray}
		\label{eq:far}
		U(x)=
      \frac{B D_R+B_R D}{2 \pi^2 \kappa_0^2 x^4}
      \;
      k_\mathrm{B}T\; 
		=-\tilde{A}\; k_\mathrm{B} T \,\frac{a^2 R^2}{x^4},
	\end{eqnarray}
	where
	\begin{equation}
		\tilde{A}  =  \frac{2 (\alpha - 1)^2  (3 \alpha + 5)}{(\alpha + 1)  (\alpha +  3)^2}.
	\end{equation}
	Note that, by taking the appropriate limits, this expression reproduces
	previous analytical results found in the literature~\cite{GBP93,LZM+11}.
	
\begin{figure}[!h]
		\centering
		\includegraphics[width=0.55\textwidth]{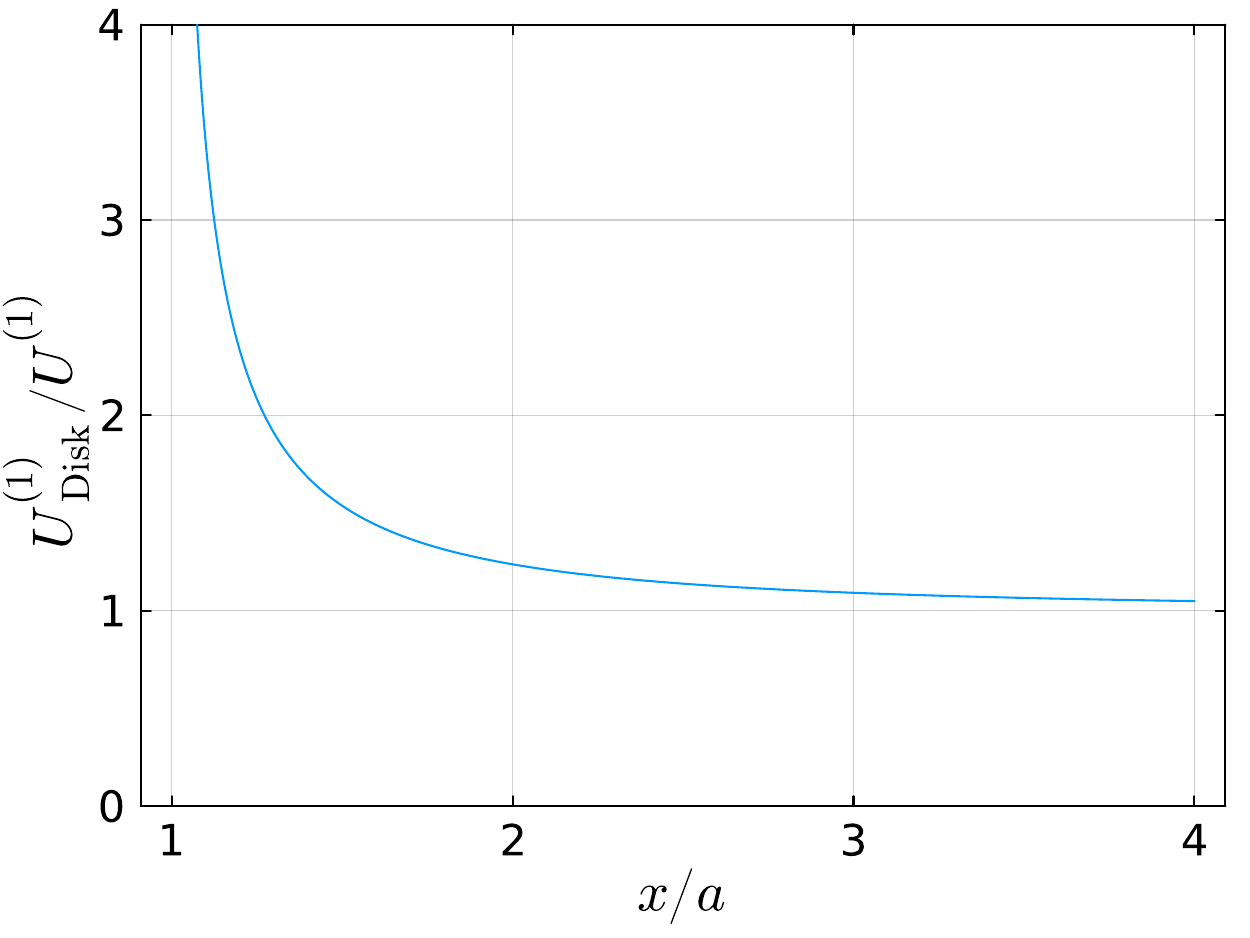}
		\caption{
         Finite-size correction to the interaction energy.
      The curve shows the ratio between the interaction energy 
      calculated 
      using
      Eq.~\ref{eq:disk_int} 
      to that from 
      Eq.~\ref{eq:general},
      with
      both
      expressions 
      expanded to first order in $(\kappa_2-\kappa_0)/\kappa_0$.
      This ratio is plotted as a function 
      of the ratio of the 
      distance
      from the wall, $x$, normalized by  
      the lateral inclusion size, $a$. 
      The correction is especially significant at short distances.
       }
		\label{fig:compare}
	\end{figure}

	When considering a single molecule diffusing in the vicinity of a sorting domain,
	one of the two regimes in Eq.~(\ref{eq:near_app}) and Eq.~(\ref{eq:far}) should be considered
	depending on the distance. A convenient
	interpolation formula for the membrane-mediated interaction energy between a
	molecule and a sorting domain of radius $R$, valid across different asymptotic
	regimes, is given by the simplest two-point Pad{\'e} approximant~\cite{GG71}
	\begin{align}
		U (r) & =  -  \hspace{0.17em} k_\mathrm{B} T \,\frac{R^2}{r^2} 
		\left[  \frac{A\,a^2}{(r - R)^2 + a^2} + (\tilde{A}-A) \frac{a^2}{r^2} \right]
		\label{eq:inter}
	\end{align}
	where $r=x+R$ is the distance from the molecule to the center of the domain.
	This  reduces to Eq.~\ref{eq:far} when $r \gg R$,  $r \gg a$,  and
	to Eq.~\ref{eq:near_app} in the limit $r \sim R$ and $r-R \gg a$, while also avoiding 
	the unphysical singularity at $x = 0$.
	\paragraph{Interaction in the proximity of the domain}
   In the previous paragraphs, we assumed that the distance $x$ between molecule
   and domain was much larger than the 
   size $a$ of the molecule.
   Here, we take into account the finite size of the inclusion by modeling 
   it as a disk of 
   rigidity~$\kappa_2$ and
   radius $a$ centered in the point of coordinates $(x,0)$: 
	\begin{equation}
		\delta\mathcal{H}_\mathrm{Disk}=
      \frac{\left(\kappa_{2}-\kappa_{0}\right)}{2}
      \int_\mathrm{Disk}\mathrm{d}^2\mathbf{r}\left\{ (\nabla^{2}u(\mathbf{r}))^{2}-2[\partial_{x}^{2}u(\mathbf{r})\partial_{y}^{2}u(\mathbf{r})-\partial_{x}\partial_{y}u(\mathbf{r})\partial_{x}\partial_{y}u(\mathbf{r})]\right\} 
	\end{equation}
	The 
   resulting 
   interaction energy can be computed as 
	\begin{eqnarray}
		\label{eq:disk_int}
		U_\mathrm{Disk} & = & -k_\mathrm{B}T\log\frac{Z}{Z_{0}}
		\;\; = \;\; -k_\mathrm{B}T\sum_{n=1}^{\infty}\frac{(-\beta)^{n}}{n!}
      \left\langle \left\{ \delta\mathcal{H}_\mathrm{Disk}
		 \right\} ^{n}\right\rangle _{0,\mathrm{c}}
	\end{eqnarray}
   where $\langle\cdots\rangle_\mathrm{c}$ denotes connected averages.
   Resummation of similar perturbation series has been performed 
   up to finite orders 
   in analogous
      geometries,
      either 
      through 
      direct computation of
      Feynman diagrams~\cite{YD12} or 
      via
      numerical 
      methods~\cite{LZM+11},
demonstrating
      that the interaction energy 
increases sharply
      at short separations.
      We have evaluated 
both 
$U$ from Eq.~\ref{eq:general}
and 
$U_\mathrm{Disk}$ from Eq.~\ref{eq:disk_int} 
      at first order  
in $(\kappa_2-\kappa_0)/\kappa_0$, getting

\begin{eqnarray*}
  \frac{U_{\mathrm{Disk}}^{(1)}}{U^{(1)}} & = & 2\left[ \frac{1}{\sqrt{1 - (a /
  x)^2}} - 1 \right]  \left( \frac{x}{a} \right)^2
\end{eqnarray*}
This relative finite-size correction
is plotted in 
Fig.~\ref{fig:compare}. 
It shows
that 
   Eq.~\ref{eq:general},
   which neglects finite-size effects, 
significantly
   underestimates 
   the interaction 
energy
   at short distances while accurately capturing its behavior at 
   separations 
   larger than the inclusion size.
	
	\section{Simulation protocol}
	\label{sec:MonteCarlo}
	
	Simulations are performed according to a protocol that employs
	a Monte Carlo technique to sample Gibbs distributed
	configurations of the membrane, and a sub-lattice continuum
	Langevin equation for particle dynamics within lattice cells.
	Each Monte Carlo sweep (MCS) is executed as follows:
	
	\paragraph*{Membrane:}
	Each site of the lattice is visited in random order, and a random
	displacement of the height of the surface at that site is proposed,
	with uniform probability
	within an interval of amplitude $2l_0$ centered
	around the previous position.
	The move is accepted or rejected
	according to the
	Metropolis criterion. The value of $l_0$ is
	chosen to achieve an acceptance
	rate of approximately $50\%$ for
	the proposed moves.
	
	\paragraph*{Diffusion:}
	After each membrane MCS,
	each lattice site $i$ is visited
	in random order. If a particle is present,  the auxiliary variables $x^{(t)}_i$ and $y^{(t)}_i$
	are updated according to the following rule:
	\begin{equation}
		\begin{split}
			x_i^{t + 1} & = x_i^t + \frac{F^t_x (x_i^t) + \sqrt{2
					\gamma k_\mathrm{B}T}\, \eta^t }{\gamma}\\
			y_i^{t + 1} & = y_i^t + \frac{F^t_y (y_i^t) + \sqrt{2
					\gamma k_\mathrm{B}T} \,\eta^t }{\gamma}
\label{eq:langevin}
		\end{split}
	\end{equation}
	where $\eta^t$ is a Gaussian noise with zero mean and variance 1, and $F^t_x(x),F^t_y(y)$ are forces acting on the molecule along the $x$ and $y$ directions at time t and position $(x,y)$. The constant $\gamma$ plays the role of the friction
	coefficient in the Langevin equation and
	sets the average length of the discrete steps of the auxiliary random walk.
	To ensure effective sampling, it is required that $\gamma \gg |F|$.
	The coordinates
	$(x_i^{(t)},y_i^{(t)})$ can be interpreted as  the
	sublattice position of the molecule at site $i$
	at time~$t$. The forces acting on the particle are evaluated as $-\nabla U$, where $U$ is the discretized membrane energy, smoothed through
	a quadratic interpolation, in order to achieve sub-lattice resolution.
	When 
reaching the jump condition
   $x_i^t>h/2$ (respectively, $<-h/2$), molecules are moved one lattice site forward (respectively, backward) along the $x$ direction. If the destination
	site is occupied, the
	molecules are not moved, and
	their position is reset to
	$x_i^t=h/2$ (respectively, $-h/2$). The same
	procedure is applied in the $y$ direction.
   When out-of-equilibrium membrane processes are simulated 
   using 
   Monte-Carlo
dynamics, setting 
the two
distinct
time-steps 
required for 
protein diffusion in the membrane plane 
and 
transverse membrane fluctuations 
is non trivial.
      This 
      issue is 
      addressed in
      Ref.~\cite{Cornet+24} (see in particular their Electronic Supplementary Information)
      and 
      relates to 
      our previous discussion about time-scale separation in Sect.~\ref{sec:phentheory}.
		To correctly describe molecular diffusion, the corresponding characteristic 
time scale 
      must be much larger than the characteristic time scale of 
      membrane fluctuations.
	In our simulations, the sublattice Langevin dynamics 
	is used to
	accurately capture
	the fast-membrane-fluctuation
	regime. 
Eq.~\ref{eq:langevin} 
shows that
the number of MCS between two consecutive jumps of a free molecule
can be estimated as $\gamma\, h^2/k_\mathrm{B}T$.
   By selecting
	a {sufficiently} large
	value of $\gamma$, we ensure that
	the particle
	samples a large-enough number of membrane configurations
   from an equilibrium distribution
	before  reaching the jump condition.
	For all the simulations performed, we set
	$\gamma=500\, k_\mathrm{B}T/h^2$.
	
	\paragraph*{Insertion:} 
	A site is randomly
	selected,
	and if it is empty,
	a particle
	is inserted
	with probability~$k_I$.
	As noted in Ref.~\cite{BAN2009}, the more rigid
	are the molecules, the lower is their diffusivity.
	In order to properly compare the results for $\bar{\rho}$ obtained
	{at} different $\kappa_1/\kappa_0$ ratios,
	it is important to ensure that,
	although $k_D$ is different for each $\kappa_1/\kappa_0$, the
	dimensionless flux $r=\phi/k_D$
	remains the same.
	This is
	accomplished
	by measuring
	the diffusion rate $k_D^{(t)}$ and the molecule density $\rho^{(t)}$
	at each MCS.
	These values are then used to
	adjust the insertion rate
	according to the formula $k_I^{(t)}=rk_D^{(t)}/(1-\rho^{(t)})$. This procedure
	guarantees
	that the
	dimensionless
	flux
	mantains the assigned value~$r$.
	Observe that since one MCS is taken as the time unit, the insertion probability $k_I$ per MCS can be interpreted as an insertion rate.
	Similarly,  the diffusion rate $k_D$ of free molecules---those
	jumping between two sites lacking occupied nearest neighbors---is
	determined as the ratio of accepted diffusive jumps.
	
	\paragraph*{Extraction:}
	If a connected component
	{containing}
	$\geq N_\mathrm{E}$
	occupied sites
	is found in the system, all
	particles in
	this connected component
	are removed.
\end{appendix}








\end{document}